\documentclass[ste,twoside]{stefano}

\usepackage{amsmath}
\usepackage{amssymb}
\usepackage{graphicx}
\usepackage{epsfig}
\usepackage{rotating}
\usepackage{cite}
\usepackage{color}

\def\makeheadbox{{%
\hbox to0pt{\vbox{\baselineskip=10dd\hrule\hbox
to\hsize{\vrule\kern3pt\vbox{\kern3pt \hbox{  {\sc Phys. Rev. D
{\bf 70}, 053010-9  (2004)} } \hbox{ {\sc
{\color{blue}{dma}}[{\color{black}{imecc}}]{\color{red}{UniCamp}}
} \hspace*{10.3cm} {\color{blue}{$\boldsymbol{\Sigma \delta
\Lambda}$}} }
\kern3pt}\hfil\kern3pt\vrule}\hrule}%
\hss}}}

\def\0{\mbox{\tiny $0$}}
\def\1{\mbox{\tiny $1$}}
\def\2{\mbox{\tiny $2$}}
\def\3{\mbox{\tiny $3$}}
\def\4{\mbox{\tiny $4$}}
\def\5{\mbox{\tiny $5$}}
\def\6{\mbox{\tiny $6$}}
\def\7{\mbox{\tiny $7$}}
\def\8{\mbox{\tiny $8$}}
\def\9{\mbox{\tiny $9$}}

\def\f14{\mbox{\tiny $\frac{1}{4}$}}
\def\infm{\mbox{\tiny $-\infty$}}
\def\infp{\mbox{\tiny $+\infty$}}
\def\MPLA#1{{\em Mod. Phys. Lett.}, {\bf A #1}}
\def\APB#1{{\em Acta Phys. Polon.}, {\bf B #1}}

\def\PLB#1{{\em Phys. Lett.}, {\bf B #1}}

\def\PRD#1{{\em Phys. Rev.}, {\bf D #1}}

\def\EPJC#1{{\em Eur. Phys. J.}, {\bf C #1}}

\def\PRP#1{{\em Phys. Rep.}, {\bf #1}}

\def\PTP#1{{\em Prog. Theor. Phys.}, {\bf #1}}

\def\IJMPA#1{{\em Int. J. Mod. Phys.}\; {\bf A #1}}

\begin{document}
%

\title{AN ANALYTIC APPROACH  TO THE WAVE PACKET FORMALISM
IN OSCILLATION PHENOMENA.}

\author{
Alex E. Bernardini\inst{1}
\and Stefano De Leo\inst{2}
}

\institute{
Department of Cosmic Rays and Chronology, State University of Campinas,\\
PO Box 6165, SP 13083-970, Campinas, Brazil,\\
{\em alexeb@ifi.unicamp.br} \and
Department of Applied Mathematics, State University of Campinas,\\
PO Box 6065, SP 13083-970, Campinas, Brazil,\\
{\em deleo@ime.unicamp.br}
}


\date{{\em April, 2004}}

\abstract{ We introduce an approximation scheme to perform an
analytic study of the oscillation phenomena in a pedagogical and
comprehensive way. By using {\em Gaussian} wave packets, we show
that the oscillation is bounded by a time-dependent vanishing
function which characterizes the {\em slippage} between the
mass-eigenstate wave packets. We also demonstrate that the wave
packet {\em spreading} represents a secondary effect which play a
significant role only in the non-relativistic limit. In our
analysis, we note the presence of a  {\em new\,} time-dependent
phase and calculate how this additional term  modifies the
oscillating character of the flavor conversion formula. Finally,
by considering {\em Box} and {\em Sine} wave packets we study how
the choice of different functions to describe the particle
localization changes the oscillation probability.}









\titlerunning{An analytic approach to the wave packet formalism}

\maketitle


\section*{I. INTRODUCTION}

Recently the great interest in the quantum oscillation phenomena
\cite{Lip95,Zra98,Beu03} has stirred up an increasing number of
works devoted to several theoretical approaches to particle mixing
and oscillations \cite{Alb03,Nir03,Giu98}. Notwithstanding the
exceptional ferment in this field, the conceptual difficulties
hidden in the oscillation formulas represent an intriguing, and
sometimes embarrassing, challenge for physicists.

The standard plane wave treatment \cite{Kay89,Kay02} is the most
elementary approach used to study the flavor oscillation problem.
However, despite being physically intuitive and simple, it is,
strictly speaking, neither rigorous nor sufficient for a complete
understanding of the physics involved in quantum oscillations. The
plane wave approach implies a perfectly well-known energy-momentum
and an infinite uncertainty on the space-time localization of the
oscillating particle. Oscillations are destroyed under these
assumptions \cite{Kay81}. In order to overcome such difficulties,
an {\em intermediate} wave packet model for ultra-relavistic
neutrinos was introduced by Kayser \cite{Kay81} and followed by
other authors \cite{Zra98,Giu91,Giu98}. Meanwhile, a common
argument against this approach is that oscillating particles are
not, and cannot be, {\em directly} observed \cite{Ric93}. It would
be more convincing to write a transition probability between
observable particles involved in the production and detection of
the oscillating particle in an {\em external} wave packet
framework \cite{Giu93,Beu03}. The particle to be studied is
represented by a relativistic propagator; it propagates between a
source and a detector, where wave packets representing the
external particles are in interaction. The function which
represents the overlap of the incoming and outgoing wave packets
in the {\em external} wave packet model corresponds to the wave
function of the propagating mass-eigenstate in the {\em
intermediate} wave packet formalism. Remarkably, it could be shown
that the probability densities for ultra-relativistic stable
oscillating particles in both frameworks are mathematically
equivalent \cite{Beu03}. Thus, it makes sense, in the {\em
external} wave packet framework, to consider a wave packet
associated with the propagating particle. However, this wave
packet picture brings up a problem, as the overlap function takes
into account not only the properties of the source, but also of
the detector. This is unusual for a wave packet interpretation and
not satisfying for causality \cite{Beu03}. This point was
clarified by Giunti \cite{Giu02B} who solves this problem by
proposing an improved version of the intermediate wave packet
model where the wave packet of the oscillating particle is
explicitly computed with field-theoretical methods in terms of
external wave packets. Despite not being applied in a completely
free way, the ({\em intermediate}) wave packet treatment commonly
simplifies the discussion of some physical aspects going with the
oscillation phenomena \cite{Giu03C,DeL04,Tak01}. In this context,
we just establish a condensed scheme to analytically study the
flavor oscillation phenomena, since, in the literature, numerous
prescriptions are somewhat confusing.

Quite generally, the analytical approaches for the mass-eigenstate
time evolution do not concern with the wave packet limitations. In
particular, {\em Gaussian} wave packets \cite{Giu98,Giu02B} enable
us to quantify the first and the second order corrections to the
oscillation character of propagating particles. In section II, we
introduce {\em Gaussian} wave packets and assume a sharply peaked
momentum distribution. Then, we approximate the mass-eigenstate
energy in order to analytically obtain the expressions for the
wave packet time evolution and for the flavor oscillation
probability. The energy expansion is taken up to the second order
term  and the wave packet {\em spreading} and {\em slippage}
effects are quantified in both non-relativistic (NR) and
ultra-relativistic (UR) propagation regimes. We also identify an
additional time-dependent phase which changes the {\em standard}
oscillating character of the flavor conversion formula.
In section III, we introduce {\em Box} and  {\em smoothly
vanishing Sine} wave packets and study how the choice of different
function in describing the particle localization could play a
significant role in the oscillation probability. We draw our
conclusions in section IV.

\section*{II. GAUSSIAN WAVE PACKETS}

The main aspects of oscillation phenomena can be understood by
studying the two flavor problem. In addition, substantial
mathematical simplifications result from the assumption that the
space dependence of wave functions is one-dimensional ($z$-axis).
Therefore, we shall use these simplifications to calculate the
oscillation probabilities. In this context, the time evolution of
flavor wave packets can be described by
\begin{eqnarray}
\Phi(z,t) &= &
\phi_{ 1}(z,t)\cos{\theta}\,\mbox{\boldmath$\nu_{ 1}$} + \phi_{ 2}(z,t)\sin{\theta}\,\mbox{\boldmath$\nu_{ 2}$}\nonumber\\
          &=& \left[\phi_{ 1}(z,t)\cos^2{\theta} + \phi_{ 2}(z,t)\sin^2{\theta}\right]\,\mbox{\boldmath$\nu_\alpha$} + \left[\phi_{ 1}(z,t) - \phi_{ 2}(z,t)\right]\cos{\theta}\sin{\theta}\,\mbox{\boldmath$\nu_\beta$}\nonumber\\
          &=& \phi_{\alpha}(z,t;\theta)\,\mbox{\boldmath$\nu_\alpha$} + \phi_{\beta}(z,t;\theta)\,\mbox{\boldmath$\nu_\beta$},
\label{0}
\end{eqnarray}
where {\boldmath$\nu_\alpha$} and {\boldmath$\nu_\beta$} are flavor-eigenstates and {\boldmath$\nu_{ 1}$} and {\boldmath$\nu_{ 2}$} are mass-eigenstates.
The probability of finding a flavor state $\mbox{\boldmath$\nu_\beta$}$ at the instant $t$ is equal to the integrated squared modulus of the $\mbox{\boldmath$\nu_\beta$}$ coefficient
\begin{eqnarray}
P(\mbox{\boldmath$\nu_\alpha$}\rightarrow\mbox{\boldmath$\nu_\beta$};t)
& = &  \int_{\infm}^{\infp}\mbox{$dz$} \,\left|\phi_{\beta}(z,t;\theta)\right|^2  =
\frac{\sin^2{[2\theta]}}{2}\left\{\, 1 - \mbox{\sc Int}(t) \, \right\},
\label{1}
\end{eqnarray}
where $\mbox{\sc Int}(t)$ represents the mass-eigenstate interference term given by
\begin{equation}
\mbox{\sc Int}(t) = Re \left[\, \int_{\infm}^{\infp}dz
\,\phi^{\dagger}_{ 1}(z,t) \phi_{ 2}(z,t) \, \right]\, .
\label{2}
\end{equation}
Let us consider mass-eigenstate wave packets given at time $t = 0$ by
\begin{equation}
\phi_s(z,0) = \left(\frac{2}{\pi a^2}\right)^{ \frac{1}{4}} \exp{\left[- \frac{z^2}{a^2}\right]} \exp{[i p_s \, z]}, ~~~~s=1,\,2.
\label{3}
\end{equation}
The wave functions which describe their time evolution are
\begin{equation}
\phi_s(z,t) = \int_{\infm}^{\infp}\frac{dp_z}{2 \pi} \,
\varphi(p_z - p_s) \exp{\left[-i\,E(p_z, m_s)\,t +i \, p_z
\,z\right]}, \label{4}
\end{equation}
where
\begin{equation}
E(p_z, m_s) = \left(p_z^2 + m_s^2\right)^{ \frac{1}{2}}~~~~\mbox{and}~~~~\varphi(p_z - p_s) =  \left(2 \pi a^2 \right)^{ \frac{1}{4}} \exp{\left[- \frac{(p_z - p_s)^2a^2}{4}\right]}.\nonumber
\end{equation}
In order to obtain the oscillation probability, we can calculate the interference term $\mbox{\sc Int}(t)$
by solving the following integral
\begin{eqnarray}
&&\int_{\infm}^{\infp}\frac{dp_z}{2 \pi} \,  \varphi(p_z - p_{ 1}) \varphi(p_z - p_{ 2})
\exp{[-i \, \Delta E(p_z) \, t]} = \nonumber\\
   && ~~~~~~~~~~~~~~~~~~~~~~\exp{\left[- \frac{(a \, \Delta{p})^2}{8}\right]}\,
\int_{\infm}^{\infp}\frac{dp_z}{2 \pi}  \, \varphi^2(p_z -
p_o)\exp{[-i \, \Delta E(p_z) \, t]}, \label{6}
\end{eqnarray}
where we have changed the $z$-integration into a $p_z$-integration
and introduced the quantities
\[ \Delta p = p_{ 1} - p_{ 2}\, ,\, \, \, \, p_o =
\frac{1}{2}(p_{ 1} + p_{ 2})\, \, \, \, \, \mbox{and} \, \, \, \,
\Delta E(p_z) = E(p_z, m_1) - E(p_z, m_2)\, . \]
The oscillation term is
bounded by the exponential function of $a \, \Delta p$ at
any instant of time. Under this condition we could never observe a
{\em pure} flavor-eigenstate. Besides, oscillations are
considerably suppressed if $a \, \Delta p > 1$. A necessary
condition to observe oscillations is that $a \, \Delta p \ll 1$.
This constraint can also be expressed by $\delta p \gg \Delta p$
where $\delta p$ is the momentum uncertainty of the particle. The
overlap between the momentum distributions is indeed relevant only
for $\delta p \gg \Delta p$. Consequently, without loss of
generality, we can assume
\begin{equation}
\mbox{\sc Int}(t) = Re \left\{\int_{\infm}^{\infp}\frac{dp_z}{2
\pi}
 \, \varphi^2(p_z - p_o)\exp{[-i \, \Delta E(p_z) \, t]} \, \right\}
\label{9}.
\end{equation}
In literature, this equation is often obtained by assuming two
mass-eigenstate wave packets described by the ``same'' momentum
distribution centered around the average momentum
$p_o$. This simplifying hypothesis also guarantees
the {\em instantaneous} creation of a {\em pure} flavor-eigenstate
{\boldmath$\nu_\alpha$} at $t = 0$ \cite{DeL04}, hence, in what follows, we shall use this simplification.

\subsection*{II.A. THE ANALYTICAL APPROACH}

In order to obtain an analytic  expression for $\phi_s(z,t)$ by
solving the integral in Eq.(\ref{4}), we firstly rewrite the
energy $E(p_z, m_s)$ as
\begin{eqnarray}
E(p_z, m_s) & = & E_s \left[1 + \frac{p_z^2 -p_o^2}{E_s^2}\right]^{ \frac{1}{2}}
             =  E_s \left[1 + \sigma_s \left(\sigma_s + 2 \mbox{v}_s\right)\right]^{ \frac{1}{2}},
\label{11}
\end{eqnarray}
where \[ E_s = (m_s^2 +p_o^2)^{ \frac{1}{2}}\, ,\, \, \, \,
\mbox{v}_s = \frac{p_o}{E_s}\, \, \, \, \mbox{and}\, \, \, \,
\sigma_s = \frac{p_z - p_o}{E_s}\,.\]
By assuming a sharply peaked
momentum distribution, i. e. $(a \, E_s)^{-1}\sim\sigma_s \ll 1$,
we can expand the energy $E(p_z, m_s)$ in a power series of
$\sigma_s$. Meanwhile, the integral in Eq.(\ref{4}) can be {\em
analytically} solved only if we consider terms up to order
$\sigma_s^2$ in the series expansion. In this case, the energy
$E(p_z, m_s)$ is approximated by
\begin{eqnarray}
E(p_z, m_s) & = & E_s \left[1 + \sigma_s \mbox{v}_s  + \frac{\sigma_s^2}{2}\left(1 - \mbox{v}_s^2 \right)\right] + \mathcal{O}(\sigma_s^3)\nonumber\\
      & \approx & E_s + p_o \sigma_s + \frac{m_s^2}{2E_s} \sigma_s^2.
\label{12}
\end{eqnarray}
The zero-order term in the previous expansion,  $E_s$, gives the
standard plane wave oscillation phase. The first-order term, $p_o
\sigma_s$,   will be responsible for the {\em slippage}
 due to the different group velocities of the
mass-eigenstate wave packets and represents a linear correction to
the standard oscillation phase \cite{DeL04}. Finally, the
second-order term, $\frac{m_s^2}{2E_s} \sigma_s^2$, which is a
(quadratic) secondary correction will give the well-known
spreading effects in the time propagation of the wave packet and
will be also responsible for a ``new'' additional phase to be
computed in the final calculation. In the case of {\em Gaussian}
momentum distributions for the mass-eigenstate wave packets, these
terms can all be {\em analytically} quantified.

\begin{figure}[h]
\begin{center}
\epsfig{file= 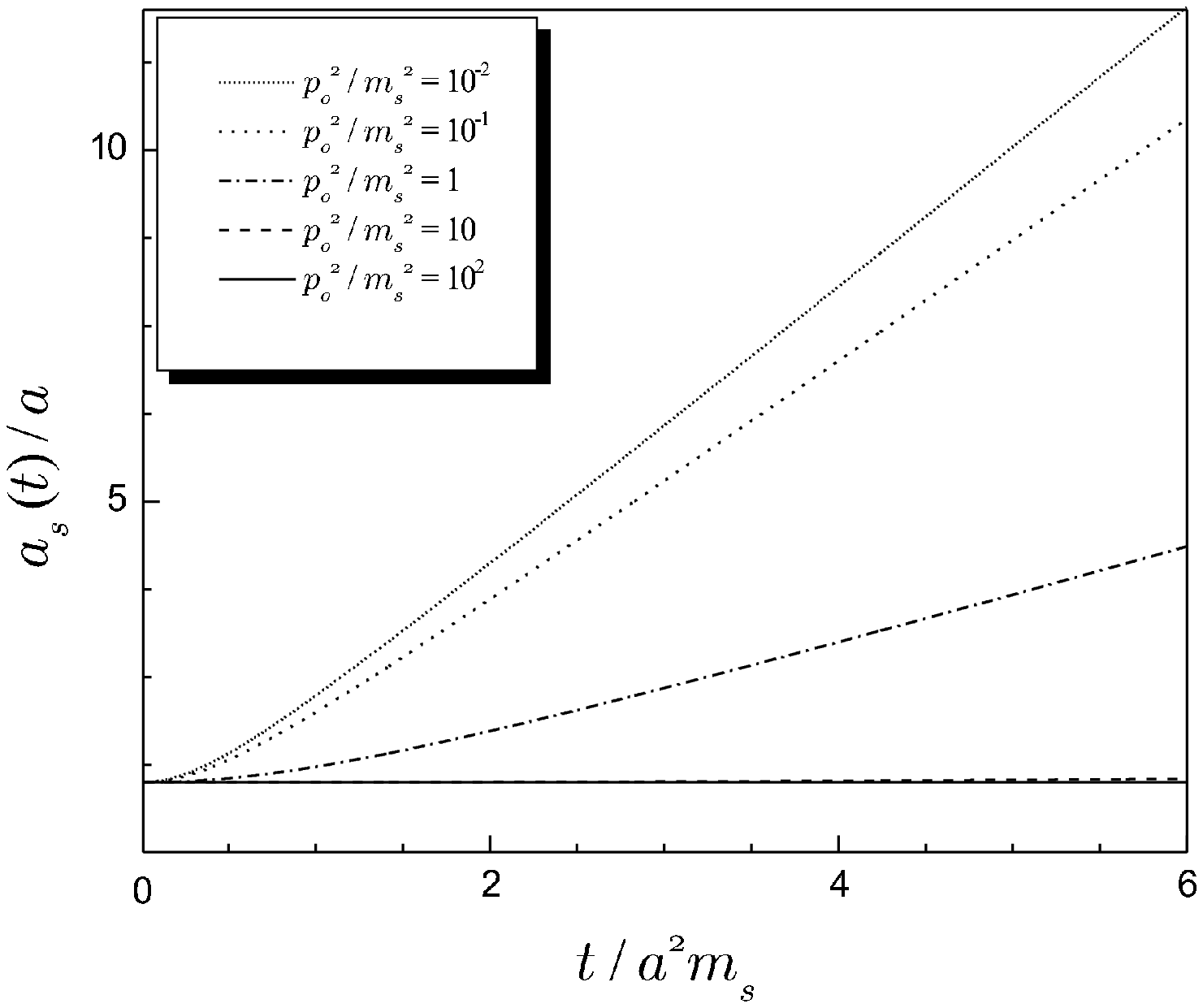, height= 13 cm, width=14.5 cm}
\end{center}
\caption{\footnotesize The time-dependence of the wave packet width $a_s(t)$ is given for different values of the ratio $p_o\, / \, m_s$.
By considering a fixed mass value $m_s$,
we compare the non-relativistic $(p_o \ll m_s)$ and the ultra-relativistic $(p_o \gg  m_s)$ propagation regimes.
We observe that the {\em spreading} is much more relevant in the former case.
In the ultra-relativistic limit $(m_s = 0)$, the wave packet does not spread and $a_s(t)$ assumes a constant value $a$.}
\label{an1}
\end{figure}

\begin{figure}[h]
\begin{center}
\epsfig{file= 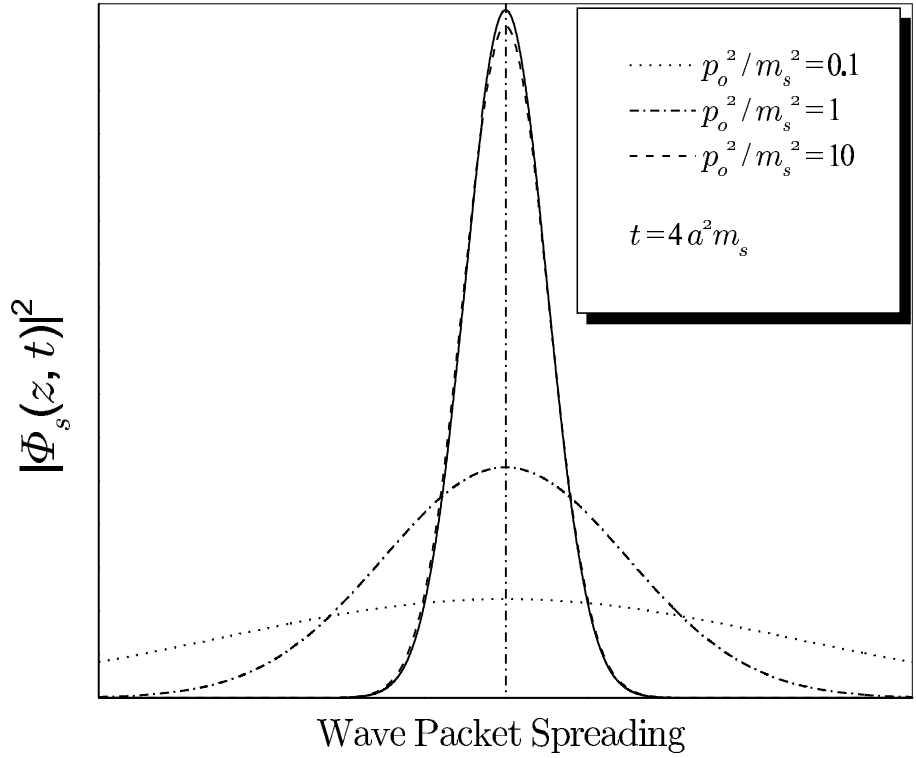, height= 10 cm, width=12 cm}
\end{center}
\caption{\footnotesize The wave packet {\em spreading} in both non-relativistic and ultra-relativistic propagation regimes
is described at time $t = 4 a^2 m_s$ in correspondence with Fig.\ref{an1}.
The solid line represent the shape of the wave packet at time $t = 0$.
In the case of an ultra-relativistic propagation expressed in terms of $\frac{p_o^2}{m_s^2} = 10$, the {\em spreading} is indeed irrelevant.}
\label{an2}
\end{figure}

By substituting (\ref{12}) in Eq.(\ref{4}) and changing the
$p_z$-integration into a $\sigma_s$-integration, we obtain the
explicit form of the mass-eigenstate wave packet time evolution,
\begin{eqnarray}
\phi_s(z,t)     & \approx &(2 \pi\, a^2)^{ \frac{1}{4}}\exp{[-i(E_s \, t - p_o \, z)]} \times\nonumber\\
&&~~~~~~~~~~~~~~~~\int_{_{-\infty}}^{^{+\infty}}\frac{d\sigma_s}{2\pi}\, E_s\,
\exp{\left[-\frac{a^2\, E_s^2 \,\sigma_s^2}{4}\right]} \exp{\left[-i\,(p_o\,t - E_s \,z)\sigma_s -i\,\frac{m_s^2 \, t}{2 E_s}\sigma_s^2\right]}\nonumber\\
                      & = &\left[\frac{2}{\pi \,a^2_s(t)}\right]^{ \frac{1}{4}}\exp{[-i\,(E_s \, t - p_o \, z)]}\,
\exp{\left[-\frac{(z - \mbox{v}_s \,t)^2}{a_s^2(t)} -i \,\theta_s(t, z)\right]},
\label{13}
\end{eqnarray}
where
\begin{equation}
~~~~a_s(t) = a \left(1 + \frac{4\, m_s^4}{a^4\, E_s^6}\,t^2\right)^{ \frac{1}{2}}
~~~~\mbox{and}~~~~
\theta_s(t, z) = \left\{\frac{1}{2}\arctan{\left[\frac{2\,m_s^2\, t}{a^2\, E_s^3}\right]} - \frac{2\, m_s^2\, t} {a^2\, E_s^3}\,\frac{(z - \mbox{v}_s \,t)^2}{a_s^2(t)}\right\}.\nonumber
\end{equation}
The time-dependent quantities $a_s(t)$ and $\theta_s(t, z)$ contain all the physically significant information which arise from the second order term in the power series expansion (\ref{12}).
The {\em spreading} of the propagating wave packet can be immediately quantified by interpreting $a_s(t)$ as a time-dependent width, i. e.
the spatial localization of the propagating particle is effectively given by $a_s(t)$ which increases during the time evolution.
In the NR propagation regime, $a_s(t)$ is approximated by $a^{\mbox{\tiny $NR$}}_s(t) = a \sqrt{1 + \frac{4}{a^4 m_s^2}t^2}$ \cite{Coh77}.
For times $t \gg a^2 m_s$ the effective wave packet width $a^{\mbox{\tiny $NR$}}_s(t)$ becomes much larger than the initial width $a$.
Otherwise, the wave packet {\em spreading} in the UR propagation regime is is approximated by $a^{\mbox{\tiny $UR$}}_s(t) = a \sqrt{1 + \frac{4 m_s^4}{a^4 p_o^6}t^2}\approx a$.
The UR {\em spreading} is practically negligible if we consider the {\em same} time-scale $T$ for both NR and UR cases, i. e. $a^{\mbox{\tiny $UR$}}_s(T) \ll a^{\mbox{\tiny $NR$}}_s(T)$.
To illustrate this characteristic, we plot the time-dependence of $a_s(t)$ in Fig.\ref{an1} where we have assumed
a particle with a definite mass value $m_s$.
By computing the squared modulus of the mass-eigenstate wave function,
\begin{equation}
|\phi_s(z,t)|^2      \approx  \left(\frac{2}{\pi a^2_s(t)}\right)^{ \frac{1}{2}}
\exp{\left[-\frac{2 (z - \mbox{v}_s \,t)^2}{a_s^2(t)}\right]},
\label{18}
\end{equation}
we illustrate the wave packet {\em spreading} in both NR and UR propagation regimes in Fig.\ref{an2} which is in correspondence with Fig.\ref{an1}.
It confirms that the wave packet {\em spreading} is irrelevant for UR particles.

Returning to Eq.(\ref{13}), we could interpret another second order effect by observing the time-behavior of the phase $\theta_s(t, z)$.
By taking into account the wave packet localization, we assume that the amplitude of the wave function is relevant in the interval $|z - \mbox{v}_s\, t| \leq a_s(t)$.
Due to the $z$-dependence,  each wave packet space-point $z$ evolves in time in a different way.
If we observe the propagation of the space-point $z = \mbox{v}_s \, t$, the crescent function $\theta_s(t,\, \mbox{v}_s t)$ assume values limited by the interval $[0, \frac{\pi}{4}[$.
Otherwise, for any other space-point given by $z = \mbox{v}_s \,t + K \,a_s(t)$, $0 < |K| \leq 1$, the phase $\theta_s(t, z)$ does not have a lower limit.
We shall show in the next subsection that the presence of a time-dependent
phase can modify the oscillation character of the flavor conversion formula.
Anyway, the phase $\theta_s(t, z)$ is not influent on the {\em free} mass-eigenstate wave packet propagation as we can see from Eq.(\ref{18}).

\subsection*{II.B. THE OSCILLATION PROBABILITY}

After having analytically quantified the second order corrections to the time evolving mass-eigenstate wave packets,
we now compute the interference term $\mbox{\sc Int}(t)$ in order to obtain an explicit expression for
the flavor conversion probability.
By solving the integral (\ref{9}) with the approximation (\ref{11})
and performing some mathematical manipulations, we obtain
\begin{equation}
\mbox{\sc Int}(t) = \mbox{\sc Bnd}(t) \times \mbox{\sc Osc}(t),
\label{20}
\end{equation}
where we have factored the time-vanishing bound of the interference term given by
\begin{equation}
\mbox{\sc Bnd}(t) = \left[1 + \mbox{\sc Sp}^2(t) \right]^{-\frac{1}{4}}
\exp{\left[-\frac{(\Delta \mbox{v} \, t)^2}{2a^2\left[1 + \mbox{\sc Sp}^2(t)\right]}\right]}
\label{21}
\end{equation}
and the time-oscillating character of the flavor conversion formula given by
\begin{equation}
\mbox{\sc Osc}(t) = Re \left\{\exp{\left[-i\Delta E \, t -i \Theta(t)\right]} \right\} = \cos{\left[\Delta E \, t + \Theta(t)\right]}
\label{22}
\end{equation}
where
\begin{equation}
\mbox{\sc Sp}(t) = \frac{t}{a^2}\Delta\left(\frac{m^2}{E^3}\right) = \rho\, \frac{\Delta \mbox{v}\, t}{a^2 \, p_o}
~~~~\mbox{and}~~~~\Theta(t) = \left[\frac{1}{2}\arctan{\left[\mbox{\sc Sp}(t)\right]} - \frac{a^2 \, p_o^2}{2 \rho^2}\frac{\mbox{\sc Sp}^3(t)}{\left[1 + \mbox{\sc Sp}^2(t)\right]}\right],
\label{24}
\end{equation}
with
\begin{equation}
\rho = 
 1 - \left[3 + \left(\frac{\Delta E}{\bar{E}}\right)^2\right] \frac{p_o^2}{\bar{E}^2}~~~~~~ \mbox{and} ~~~~~~\bar{E} = \sqrt{E_1 \, E_2}.
\label{241}
\end{equation}
The time-dependent quantities $\mbox{\sc Sp}(t)$ and $\Theta(t)$
carry the second order corrections and, consequently, the
{\em spreading} effect to the oscillation probability formula.
If $\Delta E \ll \bar{E}$, the parameter $\rho$ is limited by the interval $[1,-2]$ and it assumes the zero value when $\frac{p_o^2}{\bar{E}^2} \approx \frac{1}{3}$.
Therefore, by considering increasing values of $p_o$, from NR to UR propagation regimes,
and fixing $\frac{\Delta E}{a^2 \, \bar{E}^2}$,
the time derivatives of $\mbox{\sc Sp}(t)$ and $\Theta(t)$ have their signals inverted when $\frac{p_o^2}{\bar{E}^2}$ reaches the value $\frac{1}{3}$.

To simplify our presentation, let us study separately the time-dependent functions $\mbox{\sc Bnd}(t)$ and $\mbox{\sc Osc}(t)$.
The {\em slippage} between the mass-eigenstate wave packets is quantified by the vanishing behavior of $\mbox{\sc Bnd}(t)$.

In order to compare $\mbox{\sc Bnd}(t)$ with the correspondent function without the second order corrections (without {\em spreading}),
\begin{equation}
\mbox{\sc Bnd}_{\mbox{\tiny $WS$}}(t) = \exp{\left[-\frac{(\Delta \mbox{v} \, t)^2}{2a^2}\right]},
\label{23A}
\end{equation}
we substitute ${\mbox{\sc Sp}(t)}$ given by the expression (\ref{22}) in Eq.(\ref{21}) and we obtain the ratio
\begin{equation}
\frac{\mbox{\sc Bnd}(t)}{\mbox{\sc Bnd}_{\mbox{\tiny $WS$}}(t)} = \left[1 + \rho^2 \left(\frac{\Delta E \, t}{a^2 \, \bar{E}^2}\right)^2 \right]^{-\frac{1}{4}}
\exp{\left[\frac{\rho^2 \, p_o^2 \,\left(\Delta E \, t\right)^4}
{2\,a^6 \, \bar{E}^8\left[1 + \rho^2 \left(\frac{\Delta E \, t}{a^2 \, \bar{E}^2}\right)^2\right]}
\right]}.
\label{23}
\end{equation}
The NR limit is obtained by setting $\rho^2 = 1$ and $p_o = 0$ in Eq.(\ref{23}).
In the same way, the UR limit is obtained by setting $\rho^2 = 4$ and $p_o = \bar{E}$.

In fact, the minimal influence due to second order corrections occurs when $\frac{p_o^2}{\bar{E}^2} \approx \frac{1}{3}$ ($\rho \approx 0$).
Returning to the exponential term of Eq.(\ref{21}), we observe that the oscillation amplitude is more
relevant when $\Delta \mbox{v} \, t \ll a$.
It characterizes the {\em minimal slippage} between the mass-eigenstate wave packets which occur when the
complete spatial intersection between themselves starts to diminish during the time evolution.
Anyway, under {\em minimal slippage} conditions, we always have $\frac{\mbox{\sc Bnd}(t)}{ \mbox{\sc Bnd}_{\mbox{\tiny $WS$}}(t)} \approx 1$.

\begin{figure}[h]
\begin{center}
\epsfig{file= 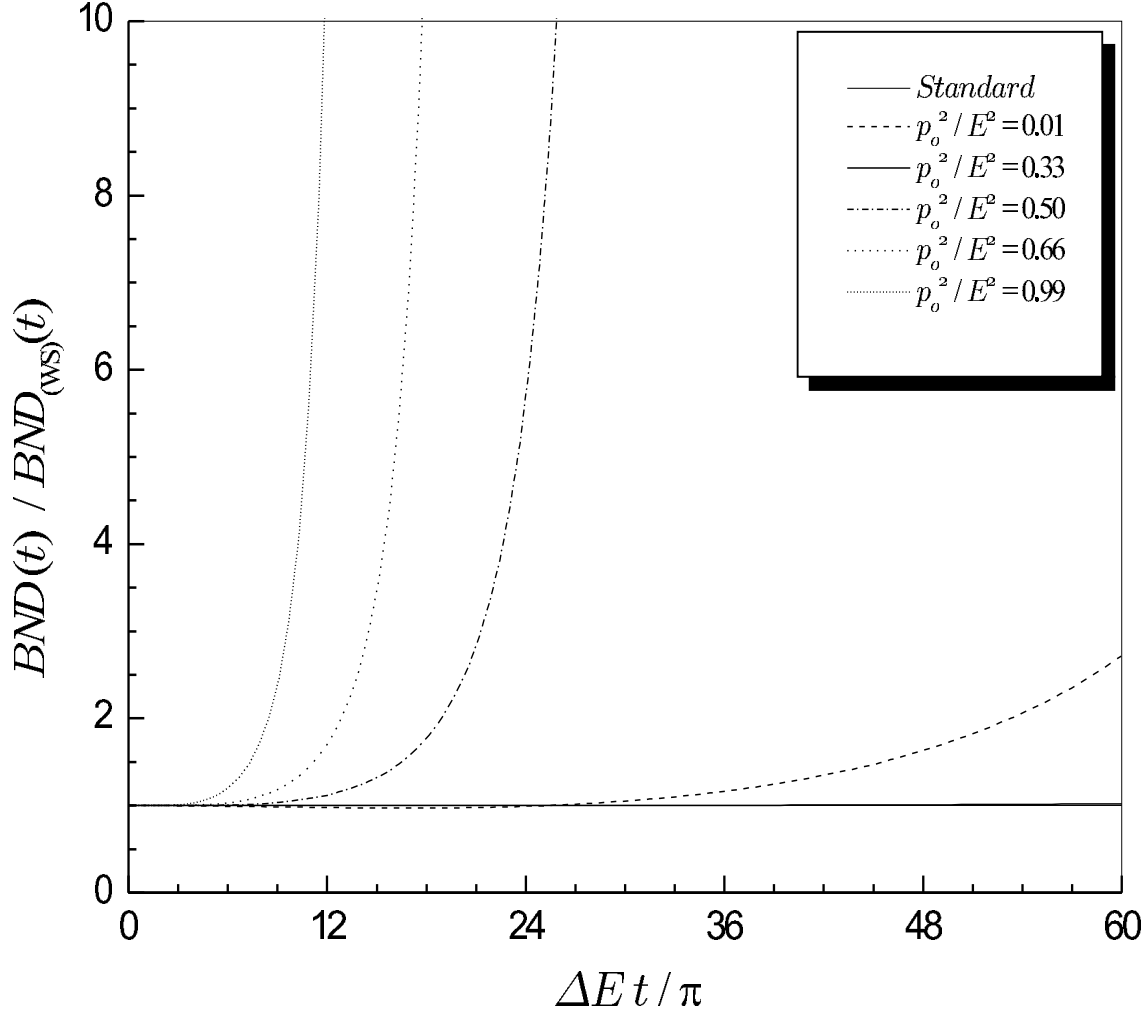, height= 13.4 cm, width=15 cm}
\end{center}
\caption{\footnotesize The comparison between the vanishing behavior {\em with} ($\mbox{\sc Bnd}(t)$) and {\em without} ($\mbox{\sc Bnd}_{\mbox{\tiny $WS$}}(t)$)
the second order corrections
for different propagation regimes.
In order to have a realistic interpretation of the information carried by the second order corrections we arbitrarily fix $a \, \bar{E} = 10$.
The second order corrections could be indeed effective for both non-relativistic and (ultra)relativistic propagation regimes,
however, the oscillations are destroyed much more rapidly in the latter case.
If $\frac{p_o^2}{\bar{E}^2} \approx \frac{1}{3}$, the second order corrections are minimal.}
\label{an3}
\end{figure}

\begin{figure}[h]
\begin{center}
\epsfig{file= 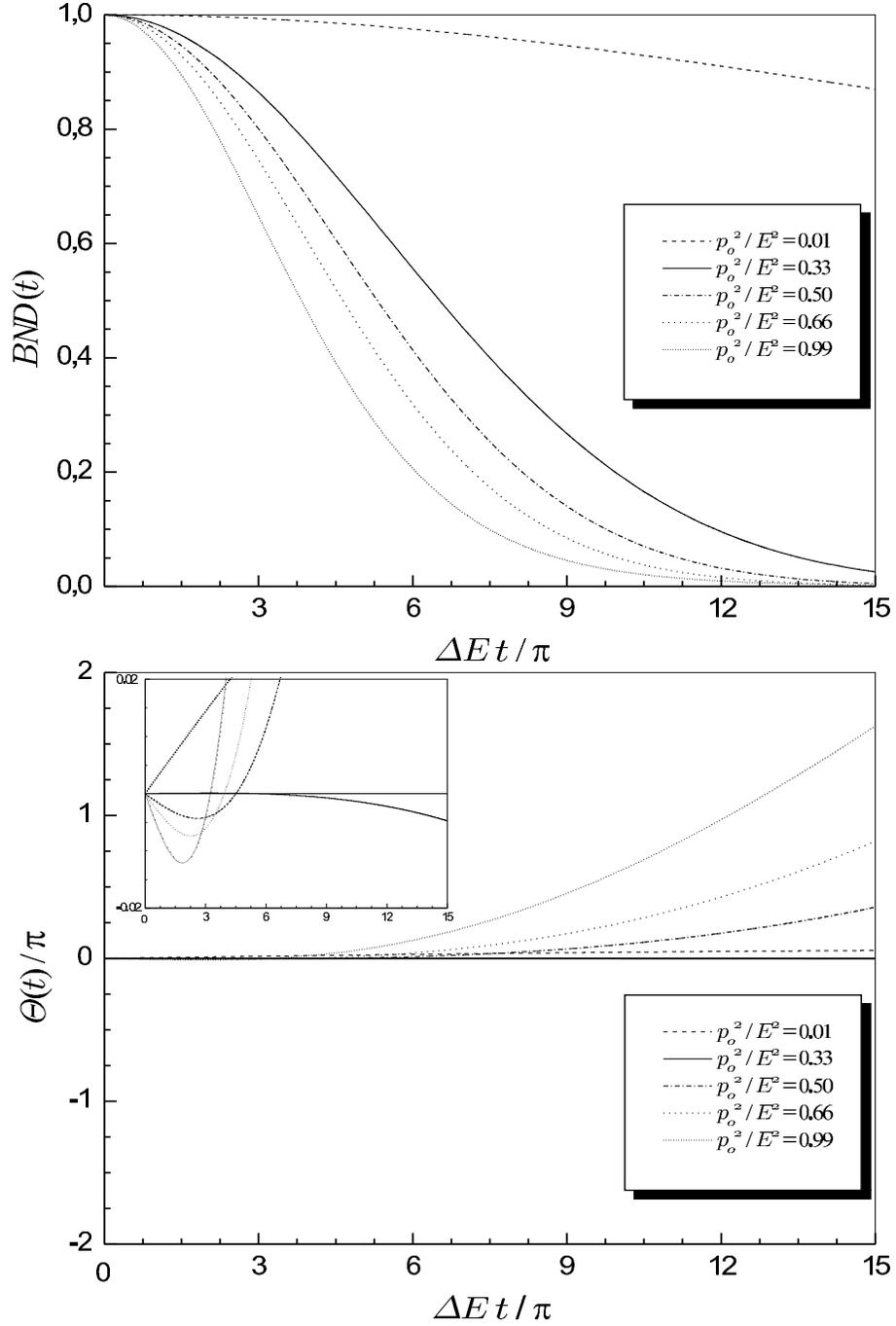, height= 18 cm, width=12 cm}
\end{center}
\caption{\footnotesize The time-behavior of the additional phase $\Theta(t)$.
The values assumed by $\Theta (t)$ are {\em  effective} while the interference term does not vanish.
In the upper box we can observe the behavior of $\mbox{\sc Bnd}(t)$ which determines the limit values effectively assumed by
$\Theta(t)$ for each propagation regime.
For relativistic regimes with $\frac{p_o^2}{\bar{E}^2} > \frac{1}{3}$, the function $\Theta(t)$ rapidly reaches its lower limit as we can observe in the small box above.
We have used $a \, \bar{E} = 10$.}
\label{an5}
\end{figure}

We plot the ratio given in Eq.(\ref{23}) for different propagation regimes in Fig.\ref{an3} where we have arbitrarily set $a\, \bar{E} = 10$.
For asymptotic times, the time-dependent term $\mbox{\sc Sp}(t)$ effectively extends the interference between the mass-eigenstate wave packets
since
\begin{equation}
\frac{\mbox{\sc Bnd}(t)}{\mbox{\sc Bnd}_{\mbox{\tiny $WS$}}(t)} \stackrel{t \rightarrow \infty}{\approx}
\frac{a \, \bar{E}} {(\rho \, \Delta E \, t)^{\frac{1}{2}} }
\exp{\left[\frac{p_o^2\left(\Delta E \, t\right)^2}{2\,a^2 \, \bar{E}^4}\right]} \gg 1,
\label{231}
\end{equation}
but, in this case, the oscillations are almost completely destroyed by $\mbox{\sc Bnd}(t)$ (see Fig.(\ref{an5})).

\begin{figure}[h]
\begin{center}
\epsfig{file= 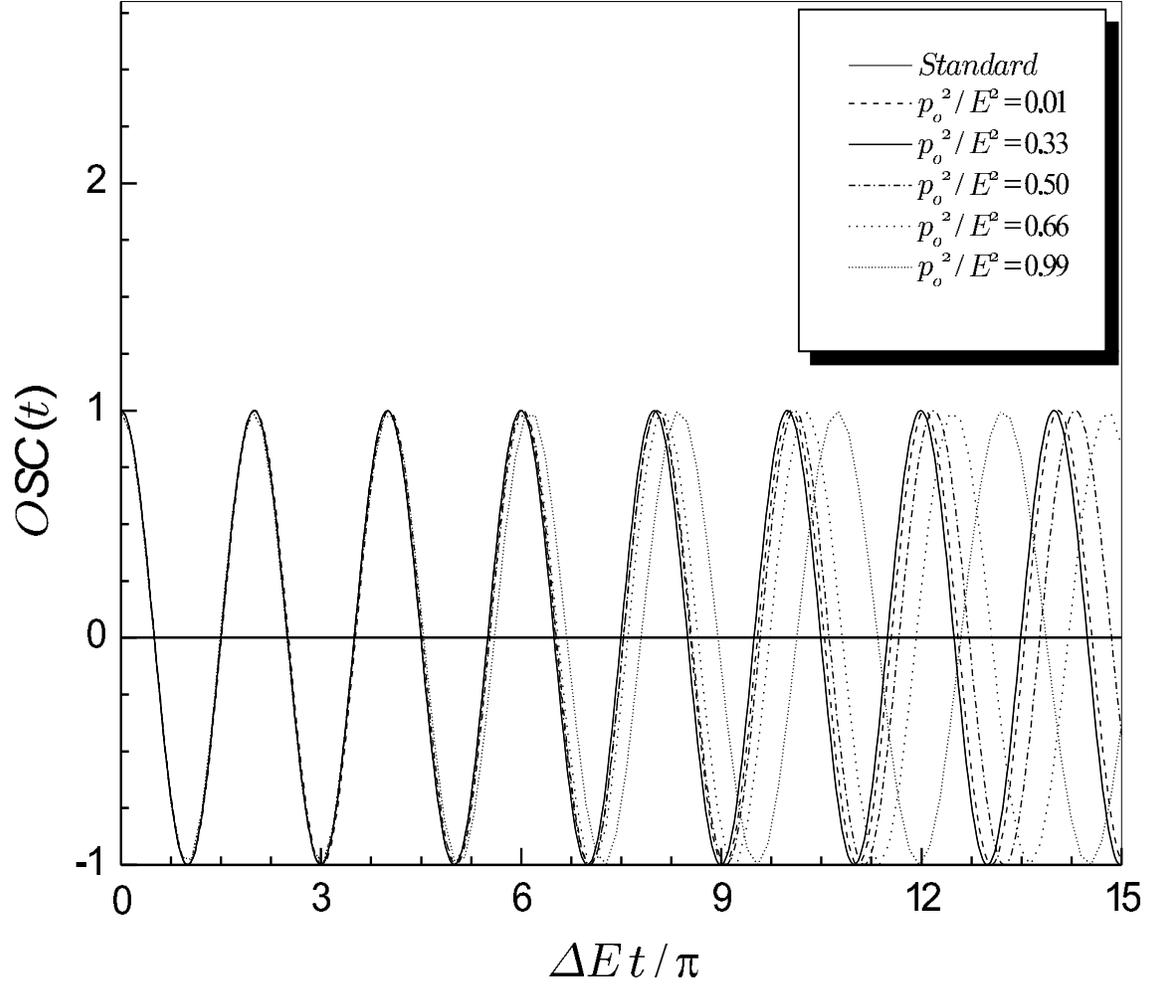, height= 13 cm, width=15 cm}
\end{center}
\caption{\footnotesize The time-behavior of $\mbox{\sc Osc}(t)$ compared with the {\em standard} plane wave oscillation given by $\cos{[\Delta E \, t]}$
for different propagation regimes.
The additional phase $\Theta(t)$ changes the oscillating character after some time of propagation.
The maximal deviation occurs for $\frac{p_o^2}{\bar{E}^2} \approx \frac{1}{3}$.}
\label{an4}
\end{figure}

The oscillating function $\mbox{\sc Osc}(t)$ of the interference term $\mbox{\sc Int}(t)$ differs from the {\em standard} oscillating term, $ \cos{[\Delta E \, t]}$,
by the presence of the additional phase $\Theta(t)$
which is essentially a second order correction.
The modifications introduced by the additional phase $\Theta(t)$ are presented in Fig.\ref{an4} where we have compared the time-behavior of $\mbox{\sc Osc}(t)$ to $\cos{[\Delta E \, t]}$ for different propagation regimes.
To study the phase $\Theta(t)$, let us conveniently define a time $t = t_o > 0$ which sets the zero of $\Theta(t)$, i. e. $\Theta(t_o) = 0$.
If $t \leq t_o$, the modulus of the phase $\Theta(t)$ reaches an upper limit when
\begin{equation}
|\Delta E \, t| = \frac{a^2 \, \bar{E}^2}{\rho \sqrt{2}}\left\{\left[\left(3 - \frac{\rho^2}{a^2\, p_o^2}\right)^2 + 4 \frac{\rho^2}{a^2\, p_o^2}\right]^{\frac{1}{2}}\, -\, \left(3 -\frac{\rho^2}{a^2\, p_o^2}\right)\right\}^{\frac{1}{2}},
\end{equation}
therefore, the maximum of $|\Theta(t)|$ depends,
not only on the propagation regime ($p_o$ value), but also on the wave packet width $a$.
Anyway, the values assumed by $|\Theta(t)|$ are restricted to the interval $[0, \frac{\pi}{4}[$.
Otherwise, if $t > t_o$, the phase $\Theta(t)$ does {\em not} have a limit and its time-dependence is essentially given by the second term of Eq.(\ref{24}).
However, it is important to notice that for $t > t_o$
the oscillating character is gradually destroyed by $\mbox{\sc Bnd}(t)$.
Consequently, another bound {\em  effective} value assumed by $\Theta (t)$
is determined by the vanishing behavior of $\mbox{\sc Bnd}(t)$.
To illustrate this point, we plot both the curves representing $\mbox{\sc Bnd}(t)$ and $\Theta(t)$ in Fig.\ref{an5} by considering the same parameters used in the study of $\mbox{\sc Bnd}(t)$.
We note the phase slowly changing in the NR regime.
The modulus of the phase $|\Theta(t)|$ rapidly reaches its upper limit when $\frac{p_o^2}{\bar{E}^2} > \frac{1}{3}$ and, after a time $t = t_o$, it continues to evolve approximately linearly in time.
But, effectively, the oscillations rapidly vanishes after $t = t_o$.

By superposing the effects of $\mbox{\sc Bnd}(t)$ and the oscillating character $\mbox{\sc Osc}(t)$ expressed in Fig.\ref{an5}, we immediately obtain the flavor oscillation probability which is explicitly given by
\begin{eqnarray}
P(\mbox{\boldmath$\nu_\alpha$}\rightarrow\mbox{\boldmath$\nu_\beta$};t)
 &\approx& \frac{\sin^2{[2\theta]}}{2}\left\{1 - \left[1 + \mbox{\sc Sp}^2(t) \right]^{-\frac{1}{4}}
\exp{\left[-\frac{(\Delta \mbox{v} \, t)^2}{2a^2\left[1 + \mbox{\sc Sp}^2(t)\right]}\right]}
\cos{\left[\Delta E \, t + \Theta(t)\right]}
  \right\}.
\label{25}
\end{eqnarray}
Obviously, the larger is the value of $a \, \bar{E}$, the smaller are the wave packet effects.
If it was sufficiently larger to not consider the second order corrections expressed in Eq.(\ref{11}),
we could  compute the oscillation probability with the leading corrections due to the {\em slippage} effect,
\begin{eqnarray}
P(\mbox{\boldmath$\nu_\alpha$}\rightarrow\mbox{\boldmath$\nu_\beta$};t)
 &\approx& \frac{\sin^2{[2\theta]}}{2}\left\{1- \exp{\left[-\frac{(\Delta \mbox{v} \, t)^2}{2\, a^2}\right]}\cos{[\Delta E \, t ]}\right\}
\label{26}
\end{eqnarray}
which corresponds to the same result obtained by \cite{DeL04}.
Under {\em minimal slippage} conditions, the above expression reproduces the {\em standard} plane wave result,
\begin{eqnarray}
P(\mbox{\boldmath$\nu_\alpha$}\rightarrow\mbox{\boldmath$\nu_\beta$};t)
 &\approx& \frac{\sin^2{[2\theta]}}{2}\left\{1- \left[1 -\frac{(\Delta \mbox{v} \, t)^2}{2\, a^2}\right]\cos{\left[\Delta E \, t \right]}\right\}\nonumber\\
 &\approx&  \frac{\sin^2{[2\theta]}}{2}\left\{1- \cos{[\Delta E \, t ]}\right\},
\label{27}
\end{eqnarray}
since we have assumed $a \, \bar{E} \gg 1$.

\section*{III. ANALYSIS WITH DIFFERENT WAVE PACKETS}

In this section we verify in what circumstances the form of the wave function can change the flavor oscillation probability.
To describe the wave packet time evolution, let us now consider a {\em Box} function and a ({\em smoothly vanishing}) {\em Sine} function in the place of a {\em Gaussian} function.
In the previous section, we have noticed it is remarkably simple to perform an analytical study with a {\em Gaussian} wave packet since its Fourier transform (FT) in the momentum space is also a {\em Gaussian} function.
In opposition, the analytical study with {\em Box} and {\em Sine} functions constrain us to perform the calculations by considering only the first order corrections in Eq. (\ref{12}), i. e.
\begin{equation}
E(p_z, m_s) \approx  E_s + p_o \sigma_s
\label{30}
\end{equation}
which only sets the {\em slippage} leading term.
We can observe from Fig.\ref{an5} that considering only the first order corrections
results in a good approximation for propagation regimes where $\frac{p_o^2}{\bar{E}^2} > \frac{1}{3}$
since the oscillations are almost completely destroyed after
any relevant second order correction takes place.
Besides, for NR propagation regimes, i e. when $\frac{p_o^2}{\bar{E}^2} < \frac{1}{3}$, by observing the Fig.\ref{an3}, we have already noticed that
the first and second order approximations are equivalent under {\em minimal slippage} conditions $\left(\frac{\mbox{\sc Bnd}(t)}{ \mbox{\sc Bnd}_{\mbox{\tiny $WS$}}(t)} \approx 1\right)$.

To simplify the discussion, we shall adopt the following definition for the initial state,
\begin{equation}
\phi^{(i)}_s(z,0) = F^{(i)}(z)\, \exp{[ip_o \, z]},
\label{31}
\end{equation}
where $i = \mbox{G}, \, \mbox{B},\,\mbox{S}$ correspond respectively to {\em Gaussian}, {\em Box} and {\em Sine} functions.
The wave packet time evolution will be expressed in terms of $\varphi^{(i)}(p_z - p_o)$ which is the FT of $\phi^{(i)}_s(z,0)$,
and the oscillation probability will be immediately computed through the expression (\ref{1}).

As we have seen in the previous section, in the case of a {\em Gaussian} function, we have
\begin{equation}
F^{({\mbox{\tiny G}})}(z) = \left(\frac{2}{\pi a^2_{\mbox{\tiny G}}}\right)^{ \frac{1}{4}} \exp{\left[- \frac{z^2}{a^2_{\mbox{\tiny G}}}\right]}~~~~\mbox{and}~~~~\varphi^{({\mbox{\tiny G}})}(p_z - p_o) = \left(2 \pi a^2_{\mbox{\tiny G}}\right)^{ \frac{1}{4}} \exp{\left[- \frac{a^2_{\mbox{\tiny G}}\,(p_z - p_o)^2 }{4}\right]}.\nonumber
\end{equation}
In this case, the wave packet has the form
\begin{eqnarray}
\phi^{(\mbox{\tiny G})}_s(z,t)  & \approx &\left(\frac{2}{\pi \,a^2_{\mbox{\tiny G}}}\right)^{ \frac{1}{4}}\exp{[-i\,(E_s \, t - p_o \, z)]}\,
\exp{\left[-\frac{(z - \mbox{v}_s \,t)^2}{a_{\mbox{\tiny G}}^2}\right]}
\label{32}
\end{eqnarray}
and the oscillation probability is reproduced by Eq.(\ref{26}).
Obviously, such results could be directly obtained by setting $a_s(t) = a$ and $\theta_s(t, z) = 0$ in Eq.(\ref{13}).

In the case of a {\em Box} function we have
\begin{equation}
F^{(\mbox{\tiny B})}(z) =\left\{\begin{array}{cll} a_{\mbox{\tiny B}}^{ -\frac{1}{2}}&& ~~~~z \in \mbox{$\left[-\frac{a_{\mbox{\tiny B}}}{2},\,\frac{a_{\mbox{\tiny B}}}{2}\right]$}\\ &&\\ 0&& ~~~~z \in\hspace{-0.3cm}\slash\hspace{0.1cm} \mbox{$\left[-\frac{a_{\mbox{\tiny B}}}{2},\,\frac{a_{\mbox{\tiny B}}}{2}\right] $}\end{array}\right. ~~~~\mbox{and}~~~~\varphi^{(\mbox{\tiny B})}(p_z - p_o) = \frac{2}{ a_{\mbox{\tiny B}}^{ \frac{1}{2}}\,(p_z - p_o)} \sin{\left[\frac{ a_{\mbox{\tiny B}}\,(p_z - p_o)}{2}\right]}.\nonumber
\end{equation}
In this case, the wave packet has the form
\begin{eqnarray}
\phi^{(\mbox{\tiny B})}_s(z,t)  & \approx & \left\{\begin{array}{cll} a_{\mbox{\tiny B}}^{ -\frac{1}{2}}\exp{[-i\,(E_s \, t - p_o \, z)]}&& ~~~~z \in \mbox{$\left[\mbox{v}_s \,t - \frac{a_{\mbox{\tiny B}}}{2},\,\mbox{v}_s \,t + \frac{a_{\mbox{\tiny B}}}{2}\right]$}\\ &&\\ 0&& ~~~~z \in\hspace{-0.3cm}\slash\hspace{0.1cm} \mbox{$\left[\mbox{v}_s \,t -\frac{a_{\mbox{\tiny B}}}{2},\,\mbox{v}_s \,t + \frac{a_{\mbox{\tiny B}}}{2}\right]$} \end{array}\right.
\label{33}
\end{eqnarray}
and the oscillation probability becomes
\begin{eqnarray}
P^{(\mbox{\tiny B})}(\mbox{\boldmath$\nu_\alpha$}\rightarrow\mbox{\boldmath$\nu_\beta$};t)
 &\approx& \left\{\begin{array}{cll} \frac{\sin^2{[2\theta]}}{2}\left\{1- \left[1 - \frac{\Delta \mbox{v} \, t}{a_{\mbox{\tiny B}}}\right]\cos{[\Delta E \, t ]}\right\}&&  ~~t \leq \frac{a_{\mbox{\tiny B}}}{\Delta \mbox{v}}\\ &&\\ 0&&~~ t > \frac{a_{\mbox{\tiny B}}}{\Delta \mbox{v}}\end{array}\right..
\label{34}
\end{eqnarray}

Finally, in the case of a {\em Sine} function we have
\begin{equation}
F^{(\mbox{\tiny S})}(z) =\left(\frac{a_{\mbox{\tiny S}}}{\pi}\right)^{ \frac{1}{2}} \frac{\sin{\left[z \, a_{\mbox{\tiny S}}^{ -1}\right]}}{z}
~~~~\mbox{and}~~~~\varphi^{(\mbox{\tiny S})}(p_z - p_o) =
\left\{\begin{array}{cll} (a_{\mbox{\tiny S}} \, \pi)^{ \frac{1}{2}}&& ~~~~a_{\mbox{\tiny S}}\,(p_z - p_o) \in \left[-1,\, 1\right]\\ &&\\ 0&& ~~~~a_{\mbox{\tiny S}}\,(p_z - p_o) \in\hspace{-0.3cm}\slash\hspace{0.1cm} \left[-1,\, 1\right] \end{array}\right..\nonumber
\end{equation}
In this case, the wave packet has the form
\begin{eqnarray}
\phi^{(\mbox{\tiny S})}_s(z,t)  & \approx &\left(\frac{a_{\mbox{\tiny S}}}{\pi}\right)^{\frac{1}{2}} \exp{[-i\,(E_s \, t - p_o \, z)]}\,
\frac{\sin{\left[ a_{\mbox{\tiny S}}^{ -1}\,(z - \mbox{v}_s \,t)\right]}}{(z - \mbox{v}_s \,t)}
\label{35}
\end{eqnarray}
and the oscillation probability becomes
\begin{eqnarray}
P^{(\mbox{\tiny S})}(\mbox{\boldmath$\nu_\alpha$}\rightarrow\mbox{\boldmath$\nu_\beta$};t)
 &\approx& \frac{\sin^2{[2\theta]}}{2}\left\{1- \left(\frac{a_{\mbox{\tiny S}}}{\Delta\mbox{v}\,t}\right)\sin{\left[\frac{\Delta\mbox{v}\,t}{a_{\mbox{\tiny S}}}\right]}
 \cos{[\Delta E \, t ]}\right\}.
\label{36}
\end{eqnarray}

\begin{figure}[h]
\begin{center}
\epsfig{file= 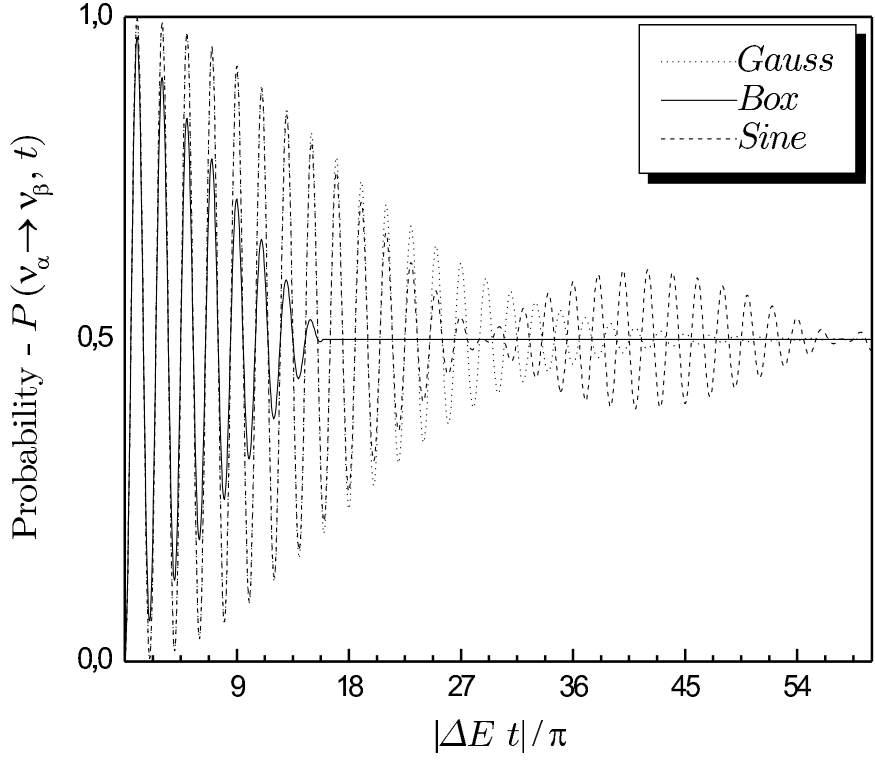, height= 10 cm, width=11.5 cm}
\end{center}
\caption{\footnotesize The flavor conversion probabilities for {\em Gaussian}, {\em Box} and {\em Sine}
wave packets by taking into account the first order correction in an analytical calculation of $\mbox{\sc Int}(t)$.
By assuming $a_{\mbox{\tiny G}}=a_{\mbox{\tiny B}} = \frac{1}{\sqrt{3}}a_{\mbox{\tiny S}}$, the {\em Gaussian} and the {\em Sine} wave packets provide
exactly the same quadratic time dependence under {\em minimal slippage} conditions whereas
the {\em Box} wave packets give a completely different behavior where the oscillation probability vanishes much more rapidly.}
\label{an6}
\end{figure}

\begin{figure}[h]
\begin{center}
\epsfig{file= 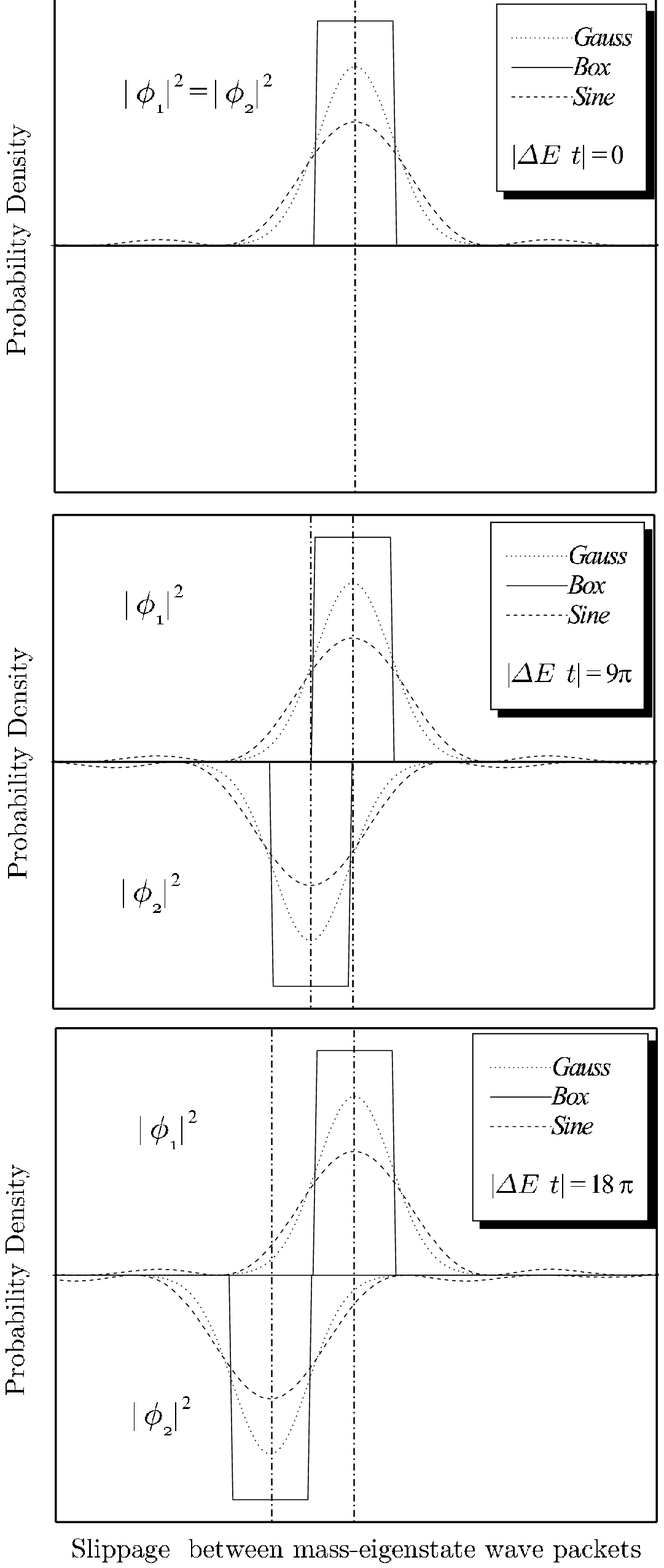, height= 17.5 cm, width=8.5cm}
\end{center}
\caption{\footnotesize The {\em slippage} between {\em Gaussian}, {\em Box} and {\em Sine} wave packets.
We can observe that the interference between the {\em Box} wave packets is abruptly interrupted
while the other two wave packets continue to interfere during longer times.
It completes the explanation of the oscillation behavior illustrated in Fig.\ref{an6}.}
\label{an7}
\end{figure}

The above results deserve some comments.
Firstly, we observe that all the three wave packet forms give the same oscillating character.
In a simplified analysis, independently of the propagation regime and without setting any parameter value,
we can compare the vanishing character of each oscillation probability
in terms of a common variable $x(t) = \frac{\Delta\mbox{v}\,t}{a_{\mbox{\tiny G}}}$.
By defining the coefficients $\alpha_{\mbox{\tiny B}} = \frac{a_{\mbox{\tiny G}}}{a_{\mbox{\tiny B}}}$
and $\alpha_{\mbox{\tiny S}} = \frac{a_{\mbox{\tiny G}}}{a_{\mbox{\tiny S}}}$ and recovering the definition of $\mbox{\sc Bnd}(t)$ we can write
\begin{equation}
\mbox{\sc Bnd}^{(\mbox{\tiny G})}(t) = \exp{\left[-\frac{x^2(t)}{2}\right]}, ~~
\mbox{\sc Bnd}^{(\mbox{\tiny B})}(t) = \left\{\begin{array}{cll} 1 - \alpha_{\mbox{\tiny B}}x(t) &&  ~ \alpha_{\mbox{\tiny B}}x(t) \leq 1 \\ &&\\ 0&&~ \alpha_{\mbox{\tiny B}}x(t) > 1\end{array}\right. ~~
\mbox{and}~~
\mbox{\sc Bnd}^{(\mbox{\tiny S})}(t) = \frac{\sin{[\alpha_{\mbox{\tiny S}}x(t)]}}{\alpha_{\mbox{\tiny S}}\,x(t)}\nonumber.
\end{equation}
Under {\em minimal slippage} conditions, i. e. when  $x(t) \ll 1$, $\mbox{\sc Bnd}^{(\mbox{\tiny G})}(t)$ and $\mbox{\sc Bnd}^{(\mbox{\tiny S})}(t)$ vanish quadratically.
Particularly, if we had set $\alpha_{\mbox{\tiny S}} = \sqrt{3}$, we would have
\begin{equation}
\mbox{\sc Bnd}^{(\mbox{\tiny G})}(t) \equiv \mbox{\sc Bnd}^{(\mbox{\tiny S})}(t) \approx 1 - \frac{x^2(t)}{2},
\end{equation}
i. e.  under {\em minimal slippage} conditions, {\em Gaussian} and {\em Sine} functions would give exactly the same oscillation probabilities.
To summarize the above results, we show the oscillation probabilities by considering the three wave packet forms in Fig.\ref{an6} where we have adopted $\alpha_{\mbox{\tiny B}} =  1$ and $\alpha_{\mbox{\tiny S}} = \sqrt{3}$.
Predominantly for {\em Sine} functions, there will always be a reminiscent oscillating character during the particle propagation.
In opposition, $\mbox{\sc Bnd}(t)^{(\mbox{\tiny B})}(t)$ vanishes linearly and the correspondent oscillation probability goes much more rapidly to zero.
Its oscillating character is suddenly ended up when $x(t) = \frac{1}{\alpha_{\mbox{\tiny B}}}$.
The {\em Sine} wave packets still provide another peculiar behavior.
Their correspondent oscillations vanish at each zero of $\sin{[x(t)]}$ but the probability returns to oscillate.
After each intermediary zero, the function $\sin{[x(t)]}$ changes the signal itself, consequently,
its maximum and minimum values are interchanged.
In Fig.\ref{an7} we illustrate the correspondent {\em slippage} between the mass-eigenstate wave packets for each case.

\section*{IV. CONCLUSIONS}

In this paper we have analytically computed the second order modifications to the flavor conversion formula
by using {\em Gaussian} wave packets.
Under the particular assumption of a sharply peaked momentum
distribution, we have obtained an explicit expression for the time evolution of the mass-eigenstates
and identified the wave packet {\em spreading}
for (U)R and NR propagation regimes.
In particular, we have observed that the {\em spreading} represents a minor
modification effect which is practically irrelevant for (ultra)relativistic propagating particles.
We have also observed the presence of an additional time-dependent phase in the
oscillating term of the flavor conversion formula.
Such an additional phase presents an analytic dependence on time
which changes the oscillating character in a peculiar way. These modifications are less relevant
when $p_o^2 \approx \frac{1}{3}\bar{E}^2$ and more relevant for NR propagation regimes
Anyway, they become completely irrelevant for UR propagatiton regimes due to the vanishing behavior
of the interference term in the oscillation probability formula.
Some influences of this additional phase on the oscillation problem were already appointed in reference \cite{Fie03}.

We know, however, that our results are strongly influenced by the {\em Gaussian} wave packet choice.
In order to understand how the wave packet form modifies the oscillation probability, we have quantified
the {\em slippage} between the mass-eigenstate
wave packets by studying a {\em Box} and a {\em Sine} localization.
In fact, by following a first order analytic approximation, a simple comparison among the different vanishing character of the oscillation probability formulas has illustrated that, under {\em minimal slippage} conditions,
the {\em Sine} and the {\em Gaussian} functions provide similar results whereas the {\em Box} function makes
the oscillations vanishing more rapidly.

To conclude, we emphasize that, an analytical study complements and clears up several aspects
already introduced in the study of quantum oscillation phenomena.

\section*{Acknowledgments}

The authors thank the University of Lecce for the hospitality and the CAPES (A.E.B) and FAEP (S.D.L) for Financial Support.

\newpage

\section*{FIGURES}


\begin{thebibliography}{99}


\bibitem{Lip95}
H. J. Lipkin, {\em Theories of nonexperiments in coherent decays of neutral mesons}, \PLB{348}, 604, (1995).

\bibitem{Zra98}
M. Zralek, {\em From kaons to neutrinos: quantum mechanics of particle oscillations}, \APB{29}, 3925 (1998).

\bibitem{Beu03}
M. Beuthe,{\em Oscillations of neutrinos and mesons in quantum field theory}, \PRP{375}, 105 (2003).

\bibitem{Alb03}
W. M. Alberico and S. M. Bilenky, {\em Neutrino oscillations, masses and mixing}, {\em arXiv:hep-ph/}0306239.

\bibitem{Nir03}
M. C. Gonzalez-Garcia and Y. Nir, {\em  Neutrino masses and mixing: evidence and implications}, {\em arXiv:hep-ph/}0202058.

\bibitem{Giu98}
C. Giunti and C. W. Kim, {\em Coherence of neutrino oscillations and wave packet approach}, \PRD{58}, 017301 (1998).

\bibitem{Kay89}
B. Kayser, F. Gibrat-Debu and F. Perrier, {\em The Physics of Massive Neutrinos}, (Cambridge University Press, Cambridge, 1989).

\bibitem{Kay02}
K. Hagiwara {\it et al.}, \PRD{66}, 010001 (2002), in [PDG Collaboration] Review of Particle Physics in {\it Neutrino physics as explored by flavor change} by B. Kayser.

\bibitem{Kay81}
B. Kayser, {\em On the quantum mechanics of neutrino oscillation}, \PRD{24}, 110 (1981).

\bibitem{Giu91}
C. Giunti C. W. Kim and U. W. Lee, {\em When do neutrinos really oscillate?: quantum mechanics of neutrino oscillations}, \PRD{44}, 3635 (1991).

\bibitem{Ric93}
J. Rich, {\em The quantum mechanics of neutrino oscillations}, \PRD{48}, 4318 (1993).





\bibitem{Giu93}
C. Giunti, C. W. Kim, J. W. Lee and U. W. Lee, {\em Treatment of neutrino oscillations without resort to weak eigenstates}, \PRD{48}, 4310 (1993).



\bibitem{Giu03C}
C. Giunti, {\em Coherence and Wave Packets in Neutrino Oscillations}, hep-th/0302026


\bibitem{DeL04}
S. De Leo, C. C. Nishi and P. Rotelli, {\em Wave Packets and Quantum Oscillations}, \IJMPA {19}, 677 (2004).

\bibitem{Tak01}
Y. Takeuchi, Y. Tazaki, S. Tsai and T. Yamazaki, {\em Wave packet approach to the equal energy/momentum/velocity prescriptions of neutrino oscillations}, \PTP{105}, 471 (2001).

\bibitem{Giu02B}
C. Giunti, {\em Neutrino wave packets in quantum field theory}, {\em JHEP} {\bf 0211}, 017 (2002).

\bibitem{Fie03}
J. Field, {\em A covariant path amplitude description of flavor oscillations:
The Gribov-Pontecorvo phase for neutrino vacuum propagation is right}, \EPJC{30}, 305  (2003).



\bibitem{Giu01}
C. Giunti, {\em Energy and momentum of oscillating neutrinos}, \MPLA{16}, 2363 (2001).

\bibitem{Coh77}
C. Cohen-Tannoudji, B. Diu and F. Laloe, {\em Quantum Mechanics} Vol.I, (John Wiley \& Sons, Paris, 1977).

\end{thebibliography}
\end{document}